\documentclass[aps,twocolumn,floatfix,superscriptaddress]{revtex4}

\usepackage[dvips]{graphics}
\usepackage{color}
\definecolor{red}{rgb}{0.85,0,0}

\begin{document}

\title{{\textcolor{red}{Logical XOR gate response in a quantum 
interferometer: A spin dependent transport}}}

\author{Moumita Dey}

\affiliation{Theoretical Condensed Matter Physics Division, Saha Institute 
of Nuclear Physics, Sector-I, Block-AF, Bidhannagar, Kolkata-700 064, India}

\author{Santanu K. Maiti}

\email{santanu.maiti@saha.ac.in}

\affiliation{Theoretical Condensed Matter Physics Division, Saha Institute 
of Nuclear Physics, Sector-I, Block-AF, Bidhannagar, Kolkata-700 064, India}

\affiliation{Department of Physics, Narasinha Dutt College, 129 Belilious 
Road, Howrah-711 101, India}

\author{S. N. Karmakar}

\affiliation{Theoretical Condensed Matter Physics Division, Saha Institute 
of Nuclear Physics, Sector-I, Block-AF, Bidhannagar, Kolkata-700 064, India}

\begin{abstract}
We examine spin dependent transport in a quantum interferometer composed of 
magnetic atomic sites based on transfer matrix formalism. The interferometer, 
threaded by a magnetic flux $\phi$, is symmetrically attached to two 
semi-infinite one-dimensional ($1$D) non-magnetic electrodes, namely, source 
and drain. A simple tight-binding model is used to describe the bridge system, 
and, here we address numerically the conductance-energy and current-voltage 
characteristics as functions of the interferometer-to-electrode coupling 
strength, magnetic flux and the orientation of local the magnetic moments 
associated with each atomic site. Quite interestingly it is observed that, 
for $\phi=\phi_0/2$ ($\phi_0=ch/e$, the elementary flux-quantum) a logical 
XOR gate like response is observed, depending on the orientation of the 
local magnetic moments associated with the magnetic atoms in the upper 
and lower arms of the interferometer, and it can be changed by an 
externally applied gate magnetic field. This aspect may be utilized in 
designing a spin based electronic logic gate.
\end{abstract}

\maketitle

\section{Introduction}

With the rapid advancement in nanoscience and nanotechnology, specially
in nanofabrication techniques~\cite{nanofab2}, study of spin 
dependent transport in mesoscopic~\cite{datta} systems has emerged as 
one of the most challenging topics in the last few decades. Analysis of 
spin transport and spin dynamics is essential to understand and develop 
the field - `spintronics'~\cite{spin1,spin4}. With the discovery of Giant 
Magneto-resistance (GMR) based magnetic field sensors~\cite{sensor} in 
1994, remarkable development has taken place in the field of magnetic 
data storage applications and quantum computation techniques. A drastic 
enhancement in computation time has been made possible using the idea of 
quantum coherence and spin entanglement. Manifestation of coherence is one 
of the most important aspect of mesoscopic systems. It is evident from 
theoretical~\cite{theoretical2,theoretical3,theoretical5,theoretical6,
theoretical7,theoretical8,oreg} and experimental~\cite{exper1,exper2,exper4} 
studies of spin transport through quantum confined 
nanostructures that the conductance of such systems depends on the spin 
state of electrons passing through the system and it can be controlled by 
an externally applied magnetic field. But measurement of current through 
these $1$D nanostructures does not reveal the feature of quantum coherence,
as it is detectable through interference experiments, most notably 
Aharonov-Bohm (AB) interferometry~\cite{gefen}. In order to study 
the effect of coherence, spin dependent transport has been studied in 
various types of ring type conductors or two path devices~\cite{chi} such 
as an AB ring or AB type interferometer with embedded 
quantum dots, with a magnetic flux $\phi$ penetrating the area enclosed 
yielding a flux dependent spin transmission probability. The study of
spin dependent transport through interferometric geometries are important 
for further development in quantum information processing as well as for 
designing spin based nano-devices. The key idea of designing spin dependent 
nano-electronic devices is based on the concept of quantum interference 
effect~\cite{imry1,bellucci1,bellucci2,peeters1,peeters2,peeters3}, and 
it is generally preserved throughout the sample having dimension smaller 
or comparable to the phase coherence length. In realistic situation, 
experimentally sizable rings are typically of the order of $0.4$-$0.6$ 
$\mu$m. Therefore, ring type conductors or two path devices are ideal 
candidates where the effect of quantum interference can be 
exploited~\cite{bohm}.

Recently, spin transport through AB type interferometers with embedded
quantum dots has drawn much attention because of its demonstration of
several physical phenomena e.g., quantum phase transitions, resonant 
tunneling and many body correlation effects. It opens a new area of 
study of spin transport, which includes spin dependent conductance 
modulation, spin filtering, spin switching, spin detecting mechanisms, 
etc. Conductance of such mesoscopic systems is associated with the 
transmission probability ($T$) of electrons, which can be calculated 
numerically by several methods like, mode matching 
techniques~\cite{modematch1}, Green's function approach~\cite{green2,
green4,san3} or transfer matrix method~\cite{transfer3,transfer4}. 

Aim of the present paper is to study the spin dependent transport through 
an AB type interferometric geometry made up of magnetic atomic sites. 
The interferometer, threaded by a magnetic flux $\phi$, is attached
symmetrically to two $1$D semi-infinite non-magnetic electrodes. A
simple tight-binding Hamiltonian is used to describe the system where 
all the calculations are done using transfer matrix formalism. Spin 
dependent conductance is calculated using the Landauer formula and also 
current-voltage characteristics are computed through the 
Landauer-B\"{u}ttiker formalism~\cite{land2,land3}. We explore 
several important features of spin transport with this simple, yet 
interesting geometry. Quite nicely we see that, at the half 
flux-quantum value of $\phi$ ($\phi=\phi_0/2$), the system exhibits
XOR gate like response depending on the orientations of the local
magnetic moments in the upper and lower arms of the interferometer, 
that can be changed by an external magnetic field. To the best of our
knowledge, the spin based XOR gate response in such a simple geometry 
has not been addressed earlier in the literature.

We organize the paper in this way. Following the introduction (Section I) 
where we address some general features and recent theoretical as well as
experimental studies on spin transport, in Section II, we describe the 
model and theoretical formulations for the calculation. In the theoretical 
formulation we describe in details the transfer matrix method following the 
renormalization procedure~\cite{anath,koiller}. Section III explores the 
numerical results, where we show the variation of conductance as functions 
of magnetic flux $(\phi)$, orientation of the local moments embedded in 
the interferometric arms and energy of the injecting electrons and then 
we illustrate the current-voltage ($I$-$V$) characteristics that clearly 
signify the logical XOR gate response. Finally, we summarize our results 
in Section IV.

\section{Model and synopsis of the theoretical background}

\begin{figure}[ht]
{\centering \resizebox*{8cm}{3.25cm}{\includegraphics{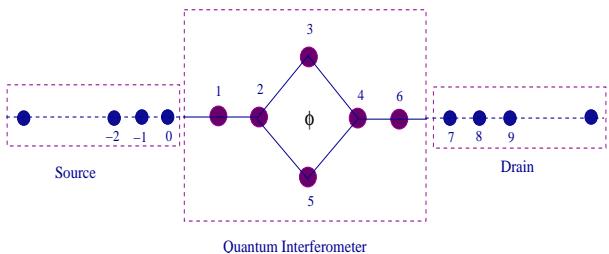}}\par}
\caption{(Color online). Schematic view of a quantum interferometer,
threaded by an AB flux $\phi$, attached to two semi-infinite $1$D 
non-magnetic electrodes, viz, source and drain. The filled purple and
blue circles correspond to the magnetic and non-magnetic atoms,
respectively.}
\label{ring}
\end{figure}

Let us begin by referring to Fig.~\ref{ring}, where a quantum 
interferometer, penetrated by an AB flux $\phi$ (measured in unit of
the elementary flux-quantum $\phi_0=ch/e$), is attached symmetrically
to two non-magnetic electrodes, viz, source and drain. The magnetic
conductor i.e., the interferometer is composed of six magnetic atoms, 
four of which are placed at four different corners of the AB type 
interferometer and the rest two are connected to the electrodes directly.
The side-attached electrodes consist of infinite number of non-magnetic 
sites labeled as $0$, $-1$, $-2$, $\ldots$, $-\infty$ for the left 
electrode (source) and $7$, $8$, $9$, $\ldots$, $\infty$ for the right 
electrode (drain). 
Each magnetic atomic site has a local magnetic moment associated with it. 
The direction of magnetization in each magnetic site is chosen to be 
arbitrary and specified by angles $\theta_n$ and $\varphi_n$ in spherical 
polar co-ordinate system for the $n$th atomic site. Here, $\theta_n$ 
represents the angle between the direction of magnetization and the 
chosen $Z$ axis, and $\varphi_n$ represents the azimuthal angle made by 
the projection of the local moment on $X$-$Y$ plane with the $X$ axis.

The Hamiltonian for the full system i.e., the 
electrode-interferometer-electrode can be described as,
\begin{equation}
H=H_D+H_L+H_R+H_{LD}+H_{DR}
\label{equ1}
\end{equation}
where, $H_D$ corresponds to the Hamiltonian of the AB type interferometric
device made up of magnetic atomic sites. $H_{L(R)}$ represents the Hamiltonian
for the left electrode i.e., source (right electrode i.e., drain), and 
$H_{LD(DR)}$ is the Hamiltonian representing the device-electrode 
coupling.

The spin polarized tight-binding Hamiltonian for the interferometer can be 
written within the non-interacting electron picture in the form,
\begin{eqnarray}
H_D & = & \sum_{n=1}^6 {\bf c_n^{\dagger} \left(\epsilon_0
-\vec{h_n}.\vec{\sigma} \right) c_n +} \sum_{i=2}^5 
{\bf \left(c_i^{\dagger}te^{i\Theta}c_{i+1}\right.} \nonumber \\
 & & {\bf \left. +~ c_{i+1}^{\dagger}te^{-i\Theta}c_i \right) +
\left(c_1^{\dagger}tc_2+c_2^{\dagger}tc_1\right) +} \nonumber \\
 & & {\bf \left(c_4^{\dagger}tc_6+c_6^{\dagger}tc_4\right)}
\label{equ2}
\end{eqnarray}
where, \\
${\bf c_n^{\dagger}}=\left(\begin{array}{cc}
c_{n \uparrow}^{\dagger} & c_{n \downarrow}^{\dagger} \end{array}\right)$;
~~ 
${\bf c_n}=\left(\begin{array}{c}
c_{n \uparrow} \\
c_{n \downarrow}\end{array}\right)$; \\
${\bf \epsilon_0}=\left(\begin{array}{cc}
\epsilon_0 & 0 \\
0 & \epsilon_0 \end{array}\right)$;
~~
${\bf t} e^{i\Theta}=te^{i\Theta}\left(\begin{array}{cc}
1 & 0 \\
0 & 1 \end{array}\right)$; \\
${\bf \vec{h_n}.\vec{\sigma}} = h_n\left(\begin{array}{cc}
\cos \theta_n & \sin \theta_n e^{-i \varphi_n} \\
\sin \theta_n e^{i \varphi_n} & -\cos \theta_n \end{array}\right)$ \\
~\\
\noindent
In Eq.~(\ref{equ2}), $1$st term corresponds to the effective on-site
energies of the interferometer. $\epsilon_0$'s are the site energies,
while the ${\bf \vec{h_n}.\vec{\sigma}}$ term represents the interaction 
of the spin (${\bf \sigma}$) of the injected electron with the local 
magnetic moment placed at the site $n$ with strength $h_n$. $\theta_n$
and $\varphi_n$ represent the orientation of the local magnetic moment 
situated at the site $n$ as mentioned earlier. This term is responsible 
for spin flip scattering at the sites. Flipping of spin violates spin 
conservation in the transport process through the magnetic conductor 
which may provide much impact in spintronic applications and we will 
discuss about it in the forthcoming sub-sections. Second term describes 
the nearest-neighbor hopping integral between the sites, at the corners 
of the interferometer, modified due to the presence of AB flux $\phi$ 
which is incorporated by the term $\Theta=2\pi \phi/4 \phi_0$. The $3$rd 
and $4$th terms represent the nearest-neighbor hopping between the atomic 
sites $1$, $2$ and $4$, $6$, respectively. 

Similarly, the Hamiltonian $H_{L(R)}$ can be expressed as,
\begin{equation}
H_{L(R)}=\sum_i {\bf c_i^{\dagger} \epsilon_{L(R)} c_i} + \sum_i
{\bf \left(c_i^{\dagger} t_{L(R)} c_{i+1} + h.c. \right)}
\label{equ3}
\end{equation}
where $\epsilon_{L(R)}$'s are the site energies of the electrodes and 
$t_{L(R)}$ is the hopping strength between the nearest-neighbor sites 
of the left (right) electrode. In this Hamiltonian, ${\bf \epsilon_{L(R)}}$
and ${\bf t_{L(R)}}$ are in the form, 
~\\
${\bf \epsilon_{L(R)}}=\left(\begin{array}{cc}
\epsilon_{L(R)} & 0 \\
0 & \epsilon_{L(R)} \end{array}\right)$
\vskip 0.2cm
\noindent
${\bf t_{L(R)}}=\left(\begin{array}{cc}
t_{L(R)} & 0 \\
0 & t_{L(R)} \end{array}\right)$ \\
~\\
In the same fashion, the conductor-electrode coupling Hamiltonian is 
described by,
\begin{equation}
H_{LD(DR)}= {\bf \left(c_{0(6)}^{\dagger} t_{LD(DR)} c_{1(7)} + 
c_{1(7)}^{\dagger} t_{LD(DR)} c_{0(6)} \right)}
\label{equ4}
\end{equation}
where, $t_{LD(DR)}$ being the device-to-electrode coupling strength.

Now, we start with the Schr\"{o}dinger equation,
\begin{equation}
H|\Phi\rangle = E|\Phi\rangle
\label{equ5}
\end{equation}
where,
\begin{equation}
|\Phi \rangle = \sum_i [ \psi_{i\uparrow} |i\uparrow\rangle +
\psi_{i\downarrow} |i\downarrow\rangle ]
\label{equ6}
\end{equation}
Here, $|\Phi\rangle$ is expressed as a linear combination of spin up 
and spin down Wannier states.

In order to calculate the spin dependent transmission probabilities 
through the interferometer, first we map the two-dimensional ($2$D) 
\begin{figure}[ht]
{\centering \resizebox*{8cm}{3cm}{\includegraphics{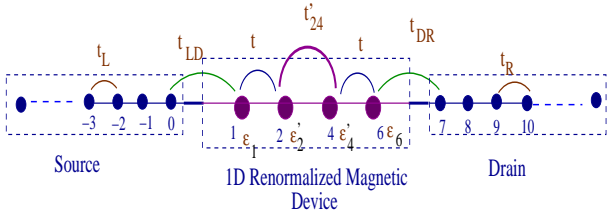}}\par}
\caption{(Color online). Schematic diagram of the renormalized $1$D
geometry. In this renormalized geometry the total number 
of atomic sites $N=4$, and, here we label the atomic sites in such a way 
(viz, $1$, $2$, $4$ and $6$) to understand the renormalized version of 
Fig.~\ref{ring} much clearly. It shows that the atomic sites $3$ and $5$
of Fig.~\ref{ring} gets renormalized.}
\label{renorm}
\end{figure}
geometry into $1$D structure by renormalization procedure. The schematic 
view of the $1$D geometry is shown in Fig.~\ref{renorm}. 

We start renormalizing the interferometric geometry by writing down 
the difference equations at the six sites of the interferometer. They 
are given as follows,
\begin{eqnarray}
\bf{(E-\epsilon_1) \psi_1} & = & \bf{t \psi_2 + t_{LD} \psi_0} \\
\bf{(E-\epsilon_2) \psi_2} & = & \bf{t \psi_1 + t_{23} \psi_3 +t_{25} 
\psi_5} \\
\bf{(E-\epsilon_3) \psi_3} & = & \bf{t_{32} \psi_2 + t_{34} \psi_4} 
\label{equ9} \\
\bf{(E-\epsilon_4) \psi_4} & = & \bf{t_{43} \psi_3 + t_{45} \psi_5 + 
t \psi_6} \\
\bf{(E-\epsilon_5) \psi_5} & = & \bf{t_{52} \psi_2 + t_{54} \psi_4}
\label{equ11} \\
\bf{(E-\epsilon_6) \psi_6} & = & \bf{t \psi_4 + t_{DR} \psi_7}
\end{eqnarray}
Here, ${\bf \epsilon_n=\left(\epsilon_0-\vec{h_n}.\vec{\sigma} \right)}$, 
$t_{23} = t_{34} = t_{45} = t_{52} = te^{i\Theta}$, and $t_{25} = t_{54} 
= t_{43} = t_{32} = te^{-i\Theta}$,\\
${\bf E}=\left(\begin{array}{cc}
E & 0 \\
0 & E \end{array}\right)$ and	
${\bf \psi_n}=\left(\begin{array}{c}
\psi_{n\uparrow} \\
\psi_{n\downarrow} \end{array}\right)$. \\
\vskip 0.01cm
\noindent
Substituting $\mathbf{\psi_3}$ and $\mathbf{\psi_5}$ using Eqs.~(\ref{equ9}) 
and (\ref{equ11}) we obtain the renormalized difference equations for 
sites $2$ and $4$ as,
\begin{equation}
\bf{(E-\epsilon_2^\prime) \psi_2}=t \psi_1 + t_{24}^\prime \psi_4
\label{13}
\end{equation}
\begin{equation}
\bf{(E-\epsilon_4^\prime) \psi_4}=t \psi_6 + t_{42}^\prime \psi_2
\label{14}
\end{equation}
Due to renormalization, the site energies of the $2$nd and $4$th sites 
get modified, and they are,
\begin{eqnarray}
{\bf \epsilon_2^{\prime}} & = & {\bf \epsilon_2+t_{23}\cdot
\left(E-\epsilon_3\right)^{-1} \cdot t_{32}} \nonumber \\
 & & {\bf + ~ t_{25}\cdot \left(E-\epsilon_5\right)^{-1} \cdot t_{52}}
\label{equ15}
\end{eqnarray}
\begin{eqnarray}
{\bf \epsilon_4^{\prime}} & = & {\bf \epsilon_4+t_{43}\cdot
\left(E-\epsilon_3\right)^{-1} \cdot t_{34}} \nonumber \\
 & & {\bf + ~ t_{45}\cdot \left(E-\epsilon_5\right)^{-1} \cdot t_{54}}
\label{equ16}
\end{eqnarray}
The hopping term between these sites ($2$ and $4$) is also modified as,
\begin{eqnarray}
{\bf t_{24}^{\prime}} & = & {\bf t_{23} \cdot
\left(E-\epsilon_3\right)^{-1} \cdot t_{34}} \nonumber \\
 & & {\bf + ~ t_{25}\cdot \left(E-\epsilon_5\right)^{-1} \cdot t_{54}}
\label{equ17}
\end{eqnarray}
and ${\bf t_{42}^{\prime}}$ is the hermitian conjugate of 
${\bf t_{24}^{\prime}}$.

With this renormalized $1$D geometry, we use transfer matrix method
to calculate spin dependent transmission probabilities ($T$) and 
current-voltage ($I$-$V$) characteristics through the bridge system.

For an arbitrary site $n$, the transfer matrix ($P$) can be defined in 
terms of the wave amplitudes of its neighboring ($n+1$) and ($n-1$) 
sites as,
\begin{equation}
\left(\begin{array}{c}
\psi_{n+1,\uparrow} \\
\psi_{n+1,\downarrow} \\
\psi_{n,\uparrow} \\
\psi_{n,\downarrow} \end{array} \right) = P
\left( \begin{array}{c}
\psi_{n,\uparrow} \\
\psi_{n,\downarrow} \\
\psi_{n-1,\uparrow} \\
\psi_{n-1,\downarrow} \end{array} \right)
\label{equ18}
\end{equation}
In our case, the transfer matrix equation for the renormalized geometry
relating the wave amplitudes at sites $0$, $-1$ and $N+1$, $N+2$ becomes,
\begin{equation}
\left(\begin{array}{c}
\psi_{N+2,\uparrow} \\
\psi_{N+2,\downarrow} \\
\psi_{N+1,\uparrow} \\
\psi_{N+1,\downarrow} \end{array} \right) = M
\left( \begin{array}{c}
\psi_{0,\uparrow} \\
\psi_{0,\downarrow} \\
\psi_{-1,\uparrow} \\
\psi_{-1,\downarrow} \end{array} \right)
\label{equ19}
\end{equation}
where, $N$ corresponds to the total number of sites in the $1$D magnetic
device after renormalization process. For our renormalized $1$D system, 
$N=4$. $M$ being the transfer matrix for the full system and it can be 
expressed as,
\begin{equation}
M=M_R \cdot P_6 \cdot P_4 \cdot P_2 \cdot P_1 \cdot M_L
\label{equ20}
\end{equation}
where, $P_1$, $P_2$, $P_4$ and $P_6$ represent the transfer matrices for
the sites labeled as $1$, $2$, $4$ and $6$, respectively. $M_L$ and $M_R$
correspond to the transfer matrices for the boundary sites at the left and
right electrodes, respectively.

To evaluate the transmission probabilities of up and down spin electrons, 
we calculate the explicit form of $M$, determining all the transfer 
matrices ($M_R$, $P_6$, $P_4$, $P_2$, $P_1$ and $M_L$). The matrices
are given below. 
\[P_{1}=
\left(\begin{array}{cccc}
\frac{E-\epsilon_0 + h_{1} \cos \theta_{1}}{t} & \frac{h_1 \sin \theta_{1} 
e^{-i \varphi_{1}}}{t} & -\frac{t_{LD}}{t} & 0  \\
\frac{h_1 \sin \theta_{1} e^{i \varphi_{1}}}{t} & \frac{E-\epsilon_0 - 
h_{1} \cos \theta_{1}}{t} & 0 & -\frac{t_{LD}}{t} \\
1 & 0 & 0 & 0 \\
0 & 1 & 0 & 0 \end{array} \right) \]
\[ P_{6}=
\left(\begin{array}{cccc}
\frac{E-\epsilon_0 + h_{6} \cos \theta_{6}}{t} & \frac{h_6 \sin \theta_{6} 
e^{-i \varphi_{6}}}{t} & -\frac{t}{t_{DR}} & 0  \\
\frac{h_6 \sin \theta_{6} e^{i \varphi_{6}}}{t} & \frac{E-\epsilon_0 - 
h_{6} \cos \theta_{6}}{t} & 0 & -\frac{t}{t_{DR}} \\
1 & 0 & 0 & 0 \\
0 & 1 & 0 & 0 \end{array} \right) \]
\[M_L=
\left(\begin{array}{cccc}
\frac{t_L}{t_{LD}} e^{i \beta_L} & 0 & 0 & 0  \\
0 & \frac{t_L}{t_{LD}} e^{i \beta_L} & 0 & 0 \\
0 & 0 & e^{i \beta_L} & 0 \\
0 & 0 & 0 & e^{i \beta_L} \end{array} \right) \]
\[M_R=
\left(\begin{array}{cccc}
e^{i \beta_R} & 0 & 0 & 0  \\
0 & e^{i \beta_R} & 0 & 0 \\
0 & 0 & \frac{t_{DR}}{t_R} e^{i \beta_L} & 0 \\
0 & 0 & 0 & \frac{t_{DR}}{t_R} e^{i \beta_L} \end{array} \right) \]
\[P_4=
\left(\begin{array}{cccc}
\frac{(E-\epsilon_4^{\prime}[1,1])}{t} & -\frac{\epsilon_4^{\prime}[1,2]}{t} & 
-\frac{t_{42}^{\prime}[1,1]}{t} & -\frac{t_{42}^{\prime}[1,2]}{t} \\
-\frac{\epsilon_4^{\prime}[2,1]}{t} & \frac{(E-\epsilon_4^{\prime}[2,2])}{t} 
& -\frac{t_{42}^{\prime}[2,1]}{t} & -\frac{t_{42}^{\prime}[2,2]}{t} \\
1 & 0 & 0 & 0 \\
0 & 1 & 0 & 0 \end{array} \right) \]
\[P_2=
\left(\begin{array}{cccc}
\frac{\alpha_1}{\alpha} & -\frac{\alpha_2}{\alpha} & 
-\frac{t t_{24}^{\prime}[2,2]}{\alpha} & \frac{t t_{24}^{\prime}[1,2]}
{\alpha} \\
-\frac{\beta_1}{\beta} & \frac{\beta_2}{\beta} & 
\frac{t t_{24}^{\prime}[2,1]}{\beta} & -\frac{t t_{24}^{\prime}[1,1]}
{\beta} \\
1 & 0 & 0 & 0 \\
0 & 1 & 0 & 0 \end{array} \right) \]
where, 
\begin{eqnarray*}
\alpha_1 & = & t_{24}^{\prime}[2,2] (E-\epsilon_2^{\prime}[1,1]) + 
t_{24}^{\prime}[1,2] \epsilon_2^{\prime}[2,1] \\
\alpha_2 & = & \epsilon_2^{\prime}[1,2] t_{24}^{\prime}[2,2] +  
t_{24}^{\prime}[1,2] (E-\epsilon_2^{\prime}[2,2]) \\
\alpha & = & t_{24}^{\prime}[1,1] t_{24}^{\prime}[2,2] - 
t_{24}^{\prime}[1,2] t_{24}^{\prime}[2,1] \\
\beta_1 & = & \epsilon_2^{\prime}[2,1] t_{24}^{\prime}[1,1] + 
t_{24}^{\prime}[2,1] (E-\epsilon_2^{\prime}[1,1]) \\
\beta_2 & = & t_{24}^{\prime}[1,1] (E-\epsilon_2^{\prime}[2,2]) +
t_{24}^{\prime}[2,1] \epsilon_2^{\prime}[1,2] \\
\beta & = & t_{24}^{\prime}[2,2]  t_{24}^{\prime}[1,1] - 
t_{24}^{\prime}[2,1] t_{24}^{\prime}[1,2]=\alpha \\
\end{eqnarray*}
In the above expressions $\epsilon^{\prime}[i,j]$ corresponds to $ij$-th
elements of the matrix $\mathbf{\epsilon^{\prime}}$. Similarly we call
the matrix elements of $\mathbf{t^{\prime}}$. 

The diagonal forms of $M_L$ and $M_R$ can be explained as follows. Due 
to translational invariance of the semi-infinite electrodes the wave 
amplitudes at the sites of the electrodes ($L$ and $R$) can be written 
in Bloch wave form,
\begin{eqnarray}
\psi_n & = & A e^{i nka} \nonumber \\
& = & A e^{i n\beta_{L(R)}}
\label{neweq1}
\end{eqnarray}
Here, $\beta_{L(R)} = ka$, $k$ is the wave vector and $a$ being the lattice 
spacing of the discrete model. For the $0$th site of the left electrode 
(L) we can write,
\begin{eqnarray}
\psi_{0\uparrow} & = & e^{i \beta_{L}} \psi_{-1\uparrow} \nonumber \\
\psi_{0\downarrow} & = & e^{i \beta_{L}} \psi_{-1\downarrow}
\label{neweq2}
\end{eqnarray}
where $\beta_L$ is defined by the following energy dispersion relation as,
\begin{equation}
E=\epsilon_L + 2t_{L}\cos \beta_{L}.
\label{neweq3}
\end{equation}
Now, the difference equation for the $0$th site (the boundary site of the 
left electrode) is in the form,
\begin{eqnarray}
(E-\epsilon_L)\psi_{0,\uparrow(\downarrow)} & = & t_L \psi_{-1,\uparrow
(\downarrow)} + t_{LD} \psi_{1,\uparrow(\downarrow)} 
\label{neweq4}
\end{eqnarray}
Thus,
\begin{eqnarray}
\psi_{1,\uparrow(\downarrow)} & = & \frac{(E-\epsilon_L)}{t_{LD}}
\psi_{0,\uparrow(\downarrow)} - \frac{t_L}{t_{LD}} \psi_{-1,
\uparrow(\downarrow)} \nonumber \\
& = & \frac{t_L}{t_{LD}} \left[ e^{i\beta_L} + e^{-i\beta_L}\right] 
\psi_{0,\uparrow(\downarrow)} \nonumber \\
& & - \frac{t_L}{t_{LD}} e^{-i\beta_L} \psi_{0,\uparrow(\downarrow)} 
\nonumber \\
& = & \frac{t_L}{t_{LD}} e^{i\beta_L} \psi_{0,\uparrow(\downarrow)}
\label{neweq5}
\end{eqnarray}
Using Eqs.~(\ref{neweq2}) and (\ref{neweq5}) we can construct $M_L$ which
shows the above diagonal form. In an exactly similar way we get the
diagonal form of $M_R$.

Let us first discuss the case of up spin incidence. Considering the whole 
system (left electrode-conductor-right electrode), the wave amplitudes 
at sites $0$, $-1$ and $N+1$, $N+2$ can be written in terms of reflection
and transmission amplitudes as,
\begin{eqnarray}
\psi_{-1\uparrow} & = & e^{-i\beta_L} + \rho^{\uparrow \uparrow}e^{i\beta_L} 
\nonumber\\
\psi_{-1\downarrow} & = & \rho^{\uparrow \downarrow}e^{i\beta_L} \nonumber \\
\psi_{0\uparrow} & = & 1+\rho^{\uparrow \uparrow} \nonumber \\
\psi_{0\downarrow} & = & \rho^{\uparrow \downarrow} 
\label{equ21}
\end{eqnarray}
and,
\begin{eqnarray}
\psi_{N+2\uparrow} = \tau^{\uparrow \uparrow}e^{i(N+2)\beta_R} \nonumber \\
\psi_{N+2\downarrow} = \tau^{\uparrow \downarrow}e^{i(N+2)\beta_R} \nonumber\\
\psi_{N+1\uparrow} = \tau^{\uparrow \uparrow}e^{i(N+1)\beta_R} \nonumber \\
\psi_{N+1\downarrow} = \tau^{\uparrow \downarrow}e^{i(N+1)\beta_R} 
\label{equ22}
\end{eqnarray}
where, $\rho^{\uparrow \uparrow}$ represents the reflection amplitude of
an up spin as an up spin, and $\rho^{\uparrow \downarrow}$ denotes the
the reflection amplitude of an up spin as a down spin. 
$\tau^{\uparrow \uparrow}$ corresponds to the transmission amplitude of
an up spin without any flipping, whereas $\tau^{\uparrow \downarrow}$ 
denotes the spin flip transmission amplitude. 
We solve the transfer matrix equation (Eq.~(\ref{equ19})) substituting 
the explicit expressions of the wave amplitudes from Eqs.~(\ref{equ21}) 
and (\ref{equ22}).

The transmission probabilities $T_{\uparrow \uparrow}$ and $T_{\uparrow 
\downarrow}$ are defined by the ratio of the transmitted flux to the
incident flux as,
\begin{eqnarray}
T_{\uparrow \uparrow} & = & \frac{t_R \sin\beta_R}{t_L \sin\beta_L} 
|\tau^{\uparrow \uparrow}|^2 \nonumber \\
T_{\uparrow \downarrow} & = & \frac{t_R \sin\beta_R}{t_L \sin\beta_L} 
|\tau^{\uparrow \downarrow}|^2 
\label{equ24}
\end{eqnarray}
Therefore, the total transmission probability of an up spin becomes,
\begin{equation}
T_\uparrow = T_{\uparrow \uparrow} + T_{\uparrow \downarrow}
\label{equ25}
\end{equation}
In a similar way, we can calculate the total transmission probability for 
the case of a down spin incidence as,
\begin{equation}
T_\downarrow = T_{\downarrow \downarrow} + T_{\downarrow \uparrow}
\label{equ26}
\end{equation}
Based on the Landauer conductance formula~\cite{datta}, the conductance 
$g_{\sigma \sigma^{\prime}}$ through the interferometer can be calculated. 
At much low temperatures and bias voltage, it can be expressed in the from,
\begin{equation}
g_{\sigma \sigma^{\prime}}=\frac{e^2}{h}T_{\sigma \sigma^{\prime}}
\label{equ27}
\end{equation}
The spin dependent current flowing through the interferometric geometry 
can be determined from the expression~\cite{datta},
\begin{equation}
I_{\sigma \sigma^{\prime}} (V)= \frac{e}{h} \int 
\limits_{-\infty}^{+\infty} 
\left(f_S-f_D\right) T_{\sigma \sigma^{\prime}}(E)~dE
\label{equ28}
\end{equation}
where, $f_{S(D)}=f\left(E-\mu_{S(D)}\right)$ gives the Fermi distribution
function of the two electrodes with the electrochemical potential 
$\mu_{S(D)}=E_F\pm eV/2$.

Due to spin flip scattering individual spin currents
($I_{\uparrow \uparrow}$ and $I_{\downarrow \downarrow}$) are no longer
constant quantities, whereas the total spin is conserved. However, we 
have defined spin currents by introducing the effect of spin-flipping. 
They are expressed as: $I_{\uparrow}=I_{\uparrow \uparrow} + 
I_{\uparrow \downarrow}$ and $I_{\downarrow}=I_{\downarrow \downarrow}+ 
I_{\downarrow \uparrow}$. Non conservation of spin inside the magnetic 
quantum interferometer gives rise to a torque known as {\em spin flip 
induced spin torque}~\cite{zhu}. We can define this spin torque through 
the relation,
\begin{equation}
\tau_{flip}=\frac{\partial \langle \vec{S}\rangle}{\partial t}=
\frac{i}{\hbar}\langle\left[\hat{H},\hat{S}\right]\rangle
\label{equ29}
\end{equation} 
where, $\hat{S}=\frac{\hbar}{2}\hat{\sigma}$. The spin torque may produce
an angular displacement in the orientations of the local magnetic moments
associated with each magnetic site.

\section{Numerical results and discussion}

Spin dependent transport properties through the magnetic conductor 
having interferometric geometry are studied in various aspects of 
interferometer-to-electrode coupling strength, AB flux $\phi$, external 
magnetic field and spin flipping. Here, we assume that the two non-magnetic 
(NM) electrodes are identical in nature. For our illustrative purposes, 
let us first mention the values of the different parameters those are 
considered for the numerical calculations. The on-site energies 
($\epsilon_0$) in the interferometer are chosen to be $0$. Magnitudes
of all the local magnetic moments ($\vec{h_n}$), associated with 
the atomic sites of the interferometer, are fixed at $0.5$ and henceforth 
we call it simply as $\vec{h}$. The hopping integral between 
\begin{figure}[ht]
{\centering \resizebox*{8cm}{8cm}{\includegraphics{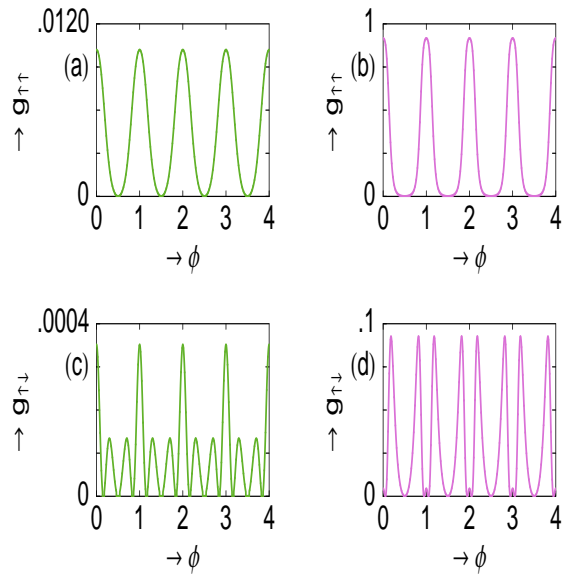}}\par}
\caption{(Color online). Typical conductance $g$ as a function of the
magnetic flux $\phi$, where the green and magenta curves correspond to the
weak- and strong-coupling limits, respectively. Other parameters are as
follows: $E=0$ and $\theta_3=\theta_5=\pi/3$.}
\label{cond1}
\end{figure}
the nearest-neighbor sites of the interferometer is set at $t=3$, while, 
for the NM electrodes it is chosen as $t_L=t_R=4$. The site energies 
($\epsilon_{L(R)}$) of all the sites in the electrodes are put to $0$. 
The azimuthal angles $\varphi_n$ for all $n$ are fixed at zero and also
the equilibrium Fermi energy $E_F$ of the conductor is set at $0$. Here, 
we choose the units $c=e=h=1$ for the sake of simplicity.

Throughout the analysis, all the essential features of spin transport 
are studied for the two distinct regimes, depending on the strength of 
coupling of the interferometer to the NM electrodes. 
\vskip 0.1cm
\noindent
{\underline{Case 1:} Weak-coupling limit.}
\vskip 0.15cm
\noindent
This regime is typically defined by the condition $t_{LD(DR)} << t$. 
Here, we choose the values of the hopping parameters as, 
$t_{LD}=t_{DR}=0.5$.
\vskip 0.1cm
\noindent
{\underline{Case 2:} Strong-coupling limit.}
\vskip 0.15cm
\noindent
This limit is described by the condition $t_{LD(DR)} \sim t$. In this 
regime, we set the values of the hopping strengths as, $t_{LD}=t_{DR}=2.5$. 

\subsection{Variation of conductance with magnetic flux $\phi$}

As representative examples, first in Fig.~\ref{cond1} we plot the 
variation of $g_{\uparrow \uparrow}$ and $g_{\uparrow \downarrow}$ for 
the interferometer as a function of magnetic flux $\phi$. The results 
are computed for the injecting electron energy $E=0$, where the green 
and magenta curves correspond to the weak- and strong-coupling limits, 
respectively. The direction of local magnetization of the magnetic atoms 
(labeled as $3$ and $5$ in Fig.~\ref{ring}) in the interferometric arms  
are chosen to be oriented at an angle $\pi/3$ with respect to the preferred 
\begin{figure}[ht]
{\centering \resizebox*{8cm}{8cm}{\includegraphics{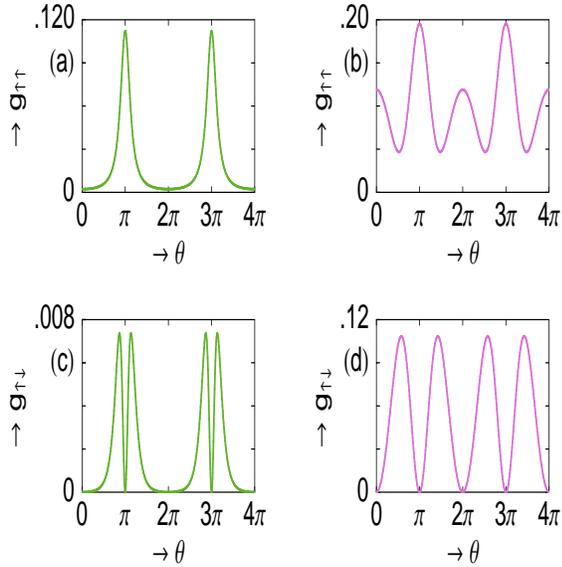}}\par}
\caption{(Color online). Typical conductance $g$ as a function of $\theta$
($\theta_3=\theta_5=\theta$), where the green and magenta curves correspond
to the weak- and strong-coupling limits, respectively. Other parameters
are as follows: $E=0$ and $\phi=\phi_0/4$.}
\label{cond2}
\end{figure}
$+Z$ direction. All the other moments are aligned along $+Z$ direction.
Figure~\ref{cond1} shows that both $g_{\uparrow \uparrow}$ and 
$g_{\uparrow \downarrow}$ vary periodically with $\phi$ showing $\phi_0$ 
flux-quantum periodicity. For a symmetrically connected interferometer 
i.e., having identical configuration in the upper and lower arms 
$(\theta_3 = \theta_5)$, the transmission probability drops exactly to 
zero at $\phi=\phi_0/2$ ($=0.5$ in our chosen unit) and it can be shown 
very easily by simple mathematical calculation as follows. 

For a symmetrically connected interferometer, the wave functions passing 
through the upper and lower arms of the interferometer are given by,
\begin{eqnarray}
\psi_1 & = & \psi_0 e^{\frac{ie}{\hbar c} \int \limits_{\gamma_1} 
\vec{A}.\vec{dr}} \nonumber \\
\psi_2 & = & \psi_0 e^{\frac{ie}{\hbar c} \int \limits_{\gamma_2} 
\vec{A}.\vec{dr}} 
\label{eqn29}
\end{eqnarray}
where, $\gamma_1$ and $\gamma_2$ are used to indicate the two different 
paths of electron propagation along the two arms of the interferometer.
$\psi_0$ denotes the wave function in absence of magnetic flux $\phi$
and it is same for both upper and lower arms as the interferometer is
symmetrically coupled to the electrodes. $\vec{A}$ is the vector
potential associated with the magnetic field $\vec{B}$ by the relation
$\vec{B}= \vec{\nabla} \times \vec{A}$. Hence the probability amplitude
of finding the electron passing through the interferometer can be 
calculated as,
\begin{equation}
|\psi_1 + \psi_2|^2 = 2|\psi_0|^2 + 2|\psi_0|^2 \cos \left({\frac{2\pi 
\phi}{\phi_0}}\right)
\label{eqn30}
\end{equation}
where, $\phi = \oint \vec{A}.\vec{dr} = \int \int \vec{B}.\vec{ds}$ 
is the flux enclosed by the interferometer.

Here, it is clearly observed from Eq.~(\ref{eqn30}) that at $\phi=\phi_0/2$,
the transmission probability of an electron exactly drops to zero. This 
\begin{figure}[ht]
{\centering \resizebox*{8cm}{8cm}{\includegraphics{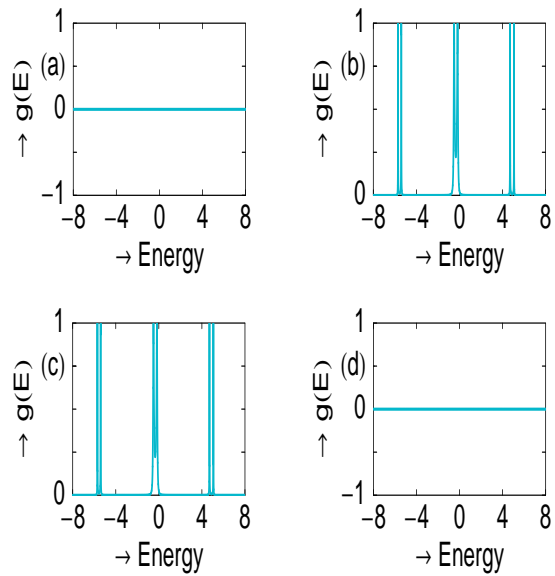}}\par}
\caption{(Color online). Up spin conductance as a function of energy $E$
for the interferometer with $\phi=\phi_0/2$ in the limit of weak-coupling.
(a) $\theta_3=\theta_5=0$, (b) $\theta_3=0$ and $\theta_5=\pi$,
(c) $\theta_3=\pi$ and $\theta_5=0$ and (d) $\theta_3=\theta_5=\pi$.}
\label{condlow}
\end{figure}
aspect can be utilized to design an XOR gate which we will describe in 
the forthcoming sub-sections. The $g$-$E$ spectra shows that the typical 
conductance gets enhanced significantly with the increase of coupling 
strength. Beside this, it is also observed that the transmission 
probability due to spin flipping is considerably smaller than the pure 
spin transmission. 

\subsection{Variation of conductance with polar angle $\theta$ of the 
magnetic moments}

In Fig.~\ref{cond2} we present the variation of $g_{\uparrow \uparrow}$ 
and $g_{\uparrow \downarrow}$ with respect to $\theta$, where $\theta$
corresponds to the angle made by the local magnetic moments in the 
interferometric arms (sites labeled as $3$ and $5$) with the preferred 
$+Z$ direction. The orientations of local magnetic moments can be changed
by applying an external magnetic field. All the other moments are aligned 
along $+$Z direction. The results are calculated for the typical magnetic 
flux $\phi=\phi_0/4$ and the injecting electron energy $E=0$. In this 
case $g_{\uparrow \uparrow}$ shows $2\pi$ periodicity as a function of 
$\theta$ both for the weak- and strong-coupling limits. But, for up and 
down orientations of the local magnetic moments in the interferometric 
arms i.e., for $\theta=0$ or any integer multiple of $\pi$, no spin 
flip takes place, and accordingly, $g_{\uparrow \downarrow}$ drops to 
zero for these configurations. 

Explanation of zero transmission probability for spin flipping is given 
as follow. Spin flip occurs due to the presence of the term 
$\vec{h}.\vec{\sigma}$ in the Hamiltonian (see Eq.~(\ref{equ2})), 
$\vec{\sigma}$ being the Pauli spin matrix with components $\sigma_x$, 
$\sigma_y$ and $\sigma_z$ for the injecting electron. The spin flipping is 
caused because of the operators $\sigma_+ (=\sigma_x + i\sigma_y)$ 
and $\sigma_- (=\sigma_x - i\sigma_y)$, respectively. For the local 
magnetic moments oriented along $\pm$ $Z$ axes, $\vec{h}.\vec{\sigma}$
$(= h_x\sigma_x + h_y\sigma_y + h_z\sigma_z)$ becomes equal to 
$h_z\sigma_z$. Accordingly, the Hamiltonian does not contain $\sigma_x$ 
and $\sigma_y$ and so as $\sigma_+$ and $\sigma_-$, which provides zero 
flipping for up or down orientation of magnetic moments. For this typical 
configuration of the localized magnetic moments mentioned above ($\pm Z$ 
axis) the spin flip torque vanishes which is clearly seen from 
Eq.~(\ref{equ29}). On the other hand, for any other orientation (apart 
from up or down configuration) of local magnetic moments a non-vanishing 
spin transfer torque appears which can provide angular displacements of 
these moments. Due to these displacements, an additional contribution 
can occur to the spin transmission which is neglected in our present 
study. 

\subsection{XOR gate response}

\subsubsection{Conductance-energy characteristics}

Let us now describe how such a simple geometric model can be implemented
as an XOR gate. For the forthcoming discussion, we set the AB flux $\phi$ 
at $\phi_0/2$.

\begin{figure}[ht]
{\centering \resizebox*{8cm}{8cm}{\includegraphics{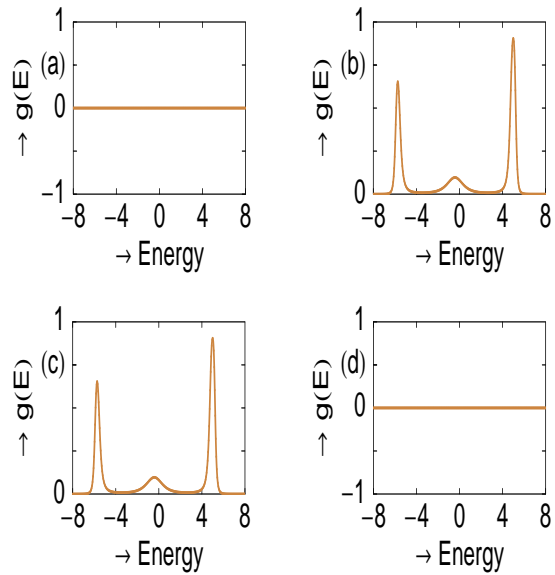}}\par}
\caption{(Color online). Up spin conductance as a function of energy $E$
for the interferometer with $\phi=\phi_0/2$ in the limit of strong-coupling.
(a) $\theta_3=\theta_5=0$, (b) $\theta_3=0$ and $\theta_5=\pi$,
(c) $\theta_3=\pi$ and $\theta_5=0$ and (d) $\theta_3=\theta_5=\pi$.}
\label{condhigh}
\end{figure}

In Figs.~\ref{condlow} and \ref{condhigh}, we show conductance-energy 
($g$-$E$) characteristics for the interferometric geometry both in the 
weak- and strong-coupling limits. With the help of external magnetic field, 
the orientation of the magnetic moments at sites $3$ and $5$ (measured 
by the parameters $\theta_3$ and $\theta_5$, respectively) can be changed. 
Here we will show that, depending on the values of $\theta_3$ and 
$\theta_5$ the interferometric geometry exhibits XOR gate response, 
keeping all the other moments oriented along $+$Z axis. These two 
($\theta_3$ and $\theta_5$) are treated as the two inputs of the XOR 
gate. Let us first discuss the case of weak-coupling (Fig.~\ref{condlow}). 
When both the two inputs to the gate are zero, i.e., $\theta_3=\theta_5=0$, 
conductance exactly drops zero (see Fig.~\ref{condlow}(a)). On the other 
hand, if any one of the two inputs is high i.e., $\theta_3$ or $\theta_5$ 
has a non-zero value, the conductance shows fine resonant peaks (see
Figs.~\ref{condlow}(b) and (c)) for some particular energy values. 
Finally, when both the inputs to the gate are high, the conductance 
again vanishes (Fig.~\ref{condlow}(d)) for the entire energy range. The 
conductance peaks are associated with the energy eigenvalues of the 
interferometer. With the increasing number of magnetic atoms, comprising 
the interferometer, number of energy levels associated with the 
interferometer increases, therefore more resonant peaks appear in the 
conductance spectrum. 

These features can be explained as follows. When both the two inputs are 
either low ($\theta_3 = \theta_5=0$) or high ($\theta_3 = \theta_5=\pi$), 
the transmission probability exactly vanishes at the half flux-quantum 
value of $\phi$ which provides zero conductance for the entire energy 
range. The vanishing behavior of conductance at $\phi=\phi_0/2$ for
these symmetric configurations of the interferometer is clearly 
understood from our earlier discussion (sub-section III$\,$A) If 
the symmetry in the orientation of the local magnetic moments in the 
two arms of the interferometer is broken by applying an external magnetic 
field i.e., $\theta_3 \neq \theta_5$ then the transmission probability 
becomes non-zero even at $\phi=\phi_0/2$. Since in this case $\theta_3$ 
and $\theta_5$ are chosen either as $0$ or $\pi$, no spin flipping takes 
place, and therefore, the contribution to $g_{\uparrow}$ comes only from 
the factor $g_{\uparrow \uparrow}$. The contribution from the spin 
flipping to the conductance spectrum will be observed for any other 
values of $\theta$ ($\theta=\theta_3=\theta_5$) apart from $0$ and 
$\pi$. Even for these orientations 
of the magnetic moments, the total conductance vanishes for the symmetric 
configuration, while it shows a finite non-zero value for the asymmetric 
one. Thus, we can conclude that the XOR gate like response will remain 
unchanged at the typical AB flux $\phi=\phi_0/2$ for any value of $\theta$.
In this interferometric geometry the asymmetry can be quantified by the 
term $\Delta \theta$ ($\Delta \theta=|\theta_3-\theta_5|$). The logical 
XOR gate response is obtained for any non-zero value of $\Delta \theta$, 
while it is best observed for the maximum value of $\Delta \theta$ 
($\Delta \theta_{max}=\pi$), which is presented in our case.

Another important thing to be mentioned here is that for $\theta=0$ or
$\pi$, spin flip transmission does not take place, and accordingly, the 
spin transfer torque $\tau_{flip}$ is zero which does not affect the 
magnetization direction, while for any other values of $\theta$ apart 
from $0$ and $\pi$, $\tau_{flip}$ is nonzero which may cause a change in 
the direction of magnetization. But the change will be the same for all 
the magnetic sites having identical magnetic moments and hence XOR gate 
feature will not get affected. Thus we can conclude that the violation
\begin{figure}[ht]
{\centering \resizebox*{8cm}{8cm}{\includegraphics{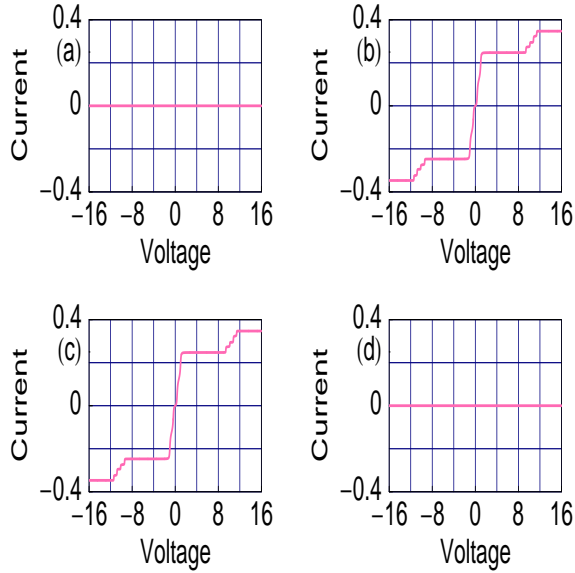}}\par}
\caption{(Color online). Up spin current $I$ as a function of applied bias
voltage $V$ for the interferometer with $\phi=\phi_0/2$ in the limit of
weak-coupling. (a) $\theta_3=\theta_5=0$, (b) $\theta_3=0$ and $\theta_5=\pi$,
(c) $\theta_3=\pi$ and $\theta_5=0$ and (d) $\theta_3=\theta_5=\pi$.}
\label{currlow}
\end{figure}
of spin conservation due to spin flip scattering should not have any 
significant impact on XOR gate response. In short, we can say that the
spin transmission probability becomes non-zero if and only if the moments 
embedded in the interferometric arms are oriented asymmetrically. Our
numerical results clearly justify the XOR gate response.

In the same footing, here we also present the conductance-energy
characteristics for the strong-coupling limit. The results are shown
in Fig.~\ref{condhigh}. All the basic features in the four different 
choices of the two inputs ($\theta_3$ and $\theta_5$) are exactly 
similar to those as presented in Fig.~\ref{condlow}, apart from the 
broadening of conductance peaks. The contribution to the broadening 
comes from the broadening of the energy levels of the interferometer 
in this strong-coupling limit. It provides a significant effect in the 
study of current-voltage ($I$-$V$) characteristics which we will describe 
in the following sub-section. It is to be noted that the conductance-energy
spectrum for down spin is exactly mirror symmetric to the spectrum 
observed for an up spin, and accordingly, we do not plot the results
further for down spin. 

\subsubsection{Current-voltage characteristics}

All the basic features of spin dependent transport obtained from 
conductance versus energy spectra can be explained in a better way
through the current-voltage ($I$-$V$) characteristics. The current 
across the quantum interferometer is computed by integrating over the 
transmission curve according to Eq.~(\ref{equ28}), where the transmission 
probability varies exactly similar to that of the conductance spectrum, 
since we get the relation $g=T$ from the Landauer conductance formula 
(Eq.~(\ref{equ27})) with $e=h=1$ in our present formulation. 

\begin{figure}[ht]
{\centering \resizebox*{8cm}{8cm}{\includegraphics{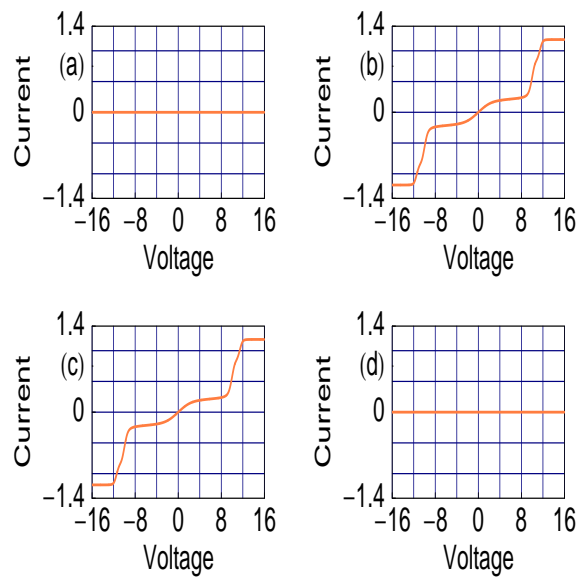}}\par}
\caption{(Color online). Up spin current $I$ as a function of applied bias
voltage $V$ for the interferometer with $\phi=\phi_0/2$ in the limit of
strong-coupling. (a) $\theta_3=\theta_5=0$, (b) $\theta_3=0$ and
$\theta_5=\pi$, (c) $\theta_3=\pi$ and $\theta_5=0$ and (d)
$\theta_3=\theta_5=\pi$.}
\label{currhigh}
\end{figure}

In Figs.~\ref{currlow} and \ref{currhigh}, we plot the current ($I$) as 
a function of applied bias voltage ($V$) both for the symmetric and 
asymmetric orientations of the 
local magnetic moments in the upper and lower arms of the interferometer. 
The flux $\phi$ is set at $\phi_0/2$ i.e., $0.5$ in our chosen unit. Let 
us first start with the case of weak-coupling (Fig.~\ref{currlow}). For 
symmetric configuration of the interferometric arms ($\theta_3=\theta_5=0$ 
and $\theta_3=\theta_5=\pi$), the current vanishes for the entire range of 
the bias voltage $V$ (Figs.~\ref{currlow}(a) and (d)). This is due to the 
fact that, for these cases the transmission probability becomes zero for 
the entire energy range as we have studied earlier. On the other hand, 
for asymmetric configuration ($\theta_3=0$, $\theta_5=\pi$ and 
$\theta_3=\pi$, $\theta_5=0$) of the interferometric arms, the current 
is non-zero (Figs.~\ref{currlow}(b) and (c)), because of non-vanishing 
spin transmission probability. This behavior becomes much more clearer 
from Table~\ref{table1}, where we make a quantitative estimate of typical 
current amplitude, computed at the bias voltage $V=10.52$. It is observed 
that, the current $I$ reaches the value to $0.294$ only when any one of 
the two inputs are high and the other is low i.e., ($\theta_3=0$ and 
$\theta_5=\pi$ or $\theta_3=\pi$ and $\theta_5=0$), while for the other 
two cases ($\theta_3=0$ and $\theta_5=0$ or $\theta_3=\pi$ and 
$\theta_5=\pi$), it ($I$) gets zero. From these $I$-$V$ characteristics 
the XOR gate like response is clearly visualized. In this weak-coupling  
limit, the current shows step-like behavior as a function of the applied 
\begin{table}[ht]
\begin{center}
\caption{XOR gate behavior in the limit of weak-coupling. The typical
current amplitude is determined at the bias voltage $V=10.52$.}
\label{table1}
~\\
\begin{tabular}{|c|c|c|}
\hline \hline
Input-I ($\theta_3$) & Input-II ($\theta_5$) & Current \em{(I)} \\ \hline
0 & 0 & 0 \\ \hline
$\pi$ & 0 & 0.294 \\ \hline
0 & $\pi$ & 0.294 \\ \hline
$\pi$ & $\pi$ & 0 \\ \hline
\end{tabular}
\end{center}
\end{table}
bias voltage $V$. This is due to the presence of sharp resonant peaks in 
the conductance spectra, as the current is obtained from integration 
method over transmission function $T$. With the increase in applied 
bias voltage $V$, the difference in chemical potentials of the two 
electrodes $(\mu_1 - \mu_2)$ increases, allowing more number of energy 
levels to fall in that range, and accordingly, more energy channels 
are accessible to the injected electrons to pass through the quantum 
\begin{table}[ht]
\begin{center}
\caption{XOR gate behavior in the limit of strong-coupling. The typical
current amplitude is determined at the bias voltage $V=10.52$.}
\label{table2}
~\\
\begin{tabular}{|c|c|c|}
\hline \hline
Input-I ($\theta_3$) & Input-II ($\theta_5$) & Current \em{(I)} \\ \hline
0 & 0 & 0 \\ \hline
$\pi$ & 0 & 0.785 \\ \hline
0 & $\pi$ & 0.785 \\ \hline
$\pi$ & $\pi$ & 0 \\ \hline
\end{tabular}
\end{center}
\end{table}
interferometer from the source to drain. Incorporation of a single 
discrete energy level i.e., a discrete quantized conduction channel,
between the range $(\mu_1 - \mu_2)$ provides a jump in the $I$-$V$ 
characteristics.  

The case of strong-coupling is depicted in Fig.~\ref{currhigh}. A 
quantitative estimate of the current is given in Table~\ref{table2}, 
where the typical current is measured for the bias voltage $V=10.52$. 
In this strong-coupling limit, all the basic features are exactly similar 
to that given in Table~\ref{table1}, only the magnitude of the output 
current gets enhanced to a value of $0.785$. The non-zero currents show 
continuum behavior with respect to the change in bias voltage. As the
sharp and discrete feature of the conductance peaks is lost in this 
strong-coupling limit, acquiring some broadening, the current changes 
continuously providing a much larger amplitude. Therefore, tuning the 
strength of the interferometer-to-electrode coupling, current can be 
enhanced significantly keeping the bias voltage constant. Exactly 
similar kind of behavior can also be observed for down spin current, 
and accordingly, we do not show the results here.

\section{Concluding remarks}

To summarize, in the present work we have explored spin dependent 
transport through interferometric geometry, penetrated by a magnetic 
flux $\phi$ using transfer matrix formalism. A simple tight-binding 
framework has been adopted to illustrate the system, where the 
interferometer comprised of magnetic atomic sites is sandwiched 
between two non-magnetic electrodes, namely, source and drain. 
We have calculated numerically the spin dependent transmission 
probability including the effect of spin-flip. Our numerical 
calculations describe conductance-energy and current-voltage 
characteristics as functions of the interferometer-to-electrode coupling 
strength, magnetic flux and the orientation of the local magnetic moments 
associated with each atom placed in interferometric arms. 

First, we have observed the variation of conductance incorporating the
effect of spin flip, as a function of magnetic flux $\phi$ showing
$\phi_0$ periodicity. It is noticed that, at half flux-quantum value
of $\phi$, transmission probability drops to zero for a symmetrically
connected interferometer. Next, we have studied the variation of
conductance with $\theta$ (considering $\theta_3=\theta_5=\theta$) for
symmetric configuration which shows that spin flipping is blocked for
$\theta=0$ or any integer multiple of $\pi$. Finally, we have obtained
XOR gate like response in $g$-$E$ and $I$-$V$ characteristics depending
on the orientations of the local magnetic moments in the upper and lower
arms of the interferometer at $\phi=\phi_0/2$. 

Throughout our work, we have addressed all the essential features
of XOR gate operation considering an interferometer with total $6$ 
atomic sites. Among them $4$ are placed at the corners to form a ring 
like structure and the rest two are coupled to the electrodes directly. 
In our model calculations, this typical number ($6$) is chosen only for 
the sake of simplicity. Though the results presented here change 
numerically with the ring size, but all the basic features remain exactly 
invariant. The main point of concern is that, whether the moments in the 
upper and lower interferometric arms are oriented symmetrically or not. 
The local magnetization direction can be changed by a rotation of the 
exchange field on the magnetic sites~\cite{theoretical8,gu}. Change of 
the local moment orientations
at sites $1$, $2$, $4$ and $6$ in our present geometric model does not
make any difference to the physical features of the results shown above.  
To be more specific, it is important to note that, in real situation the 
experimentally achievable rings have typical diameters within the range 
$0.4$-$0.6$ $\mu$m. In such a small ring, unrealistically very high 
magnetic fields are required to produce a quantum flux. To overcome this 
situation, Hod {\em et al.} have studied extensively and proposed how to 
construct nanometer scale devices, based on Aharonov-Bohm interferometry, 
those can be operated in moderate magnetic fields~\cite{baer6}.

In this work, we have calculated all the results by ignoring the effects
of temperature, electron-electron correlation, electron-phonon interaction,
disorder, etc. Here we fix the temperature at $0$K, but the basic features
will not change significantly even in non-zero finite (low) temperature
region as long as thermal energy ($k_BT$) is less than the average energy 
spacing of the energy levels of the quantum interferometer. Over the last 
few years a lot of efforts are made to 
incorporate the effect of electron-electron correlation in the study 
of spin dependent transport. Electronic correlation may cause decoherence
among the waves passing through the interferometric arms. But, at low 
temperatures the decoherence produced by electron-electron correlation
can be limited and in a very recent work Montambaux {\em et 
al.}~\cite{mont} have justified it by studying electron transport for 
some arrays of connected mesoscopic metallic rings in presence of
electronic correlation. The presence of electron-phonon interaction in
Aharonov-Bohm interferometers provides phase shifts of the conducting
electrons and due to this dephasing process electron transport through an
AB interferometer becomes highly sensitive to the AB flux $\phi$ with the
increase of electron-phonon coupling strength~\cite{hod6}. In the present
work, we have addressed our results considering the site energies of all
the atomic sites of the interferometer are identical i.e., we have treated
the ordered system. But in real case, the presence of impurities will
destroy phase coherence significantly which affects the transport 
properties. In this model it is also assumed that the two side-attached 
non-magnetic electrodes have negligible resistance. At 
the end, we would like to mention that we need further study in such 
systems by incorporating all these effects.

At the end, here we have designed a spin XOR gate using a quantum 
interferometer, based on the effect of quantum interference, which 
is a classical logic gate. On the other hand, quantum logic gates 
using ring geometries have already been proposed earlier which can 
be available in the reference~\cite{peeters}.

All these predicted results may be utilized in designing tailor made
spintronic circuits.

\end{document}